\magnification\magstep1

\centerline{\bf Quantum information processing:} \centerline{\bf
The case of vanishing interaction energy}

\medskip

\centerline{Miroljub Dugi\'c$^{1, *}$ and Milan M.\
\'Cirkovi\'c$^{2, **}$}

\medskip

\centerline{\it $^*$Department of Physics, Faculty of Science}

\centerline{\it P.~O.~Box 60, 34000 Kragujevac, Serbia}

\centerline{\it $^{**}$Astronomical Observatory, Volgina 7,
Belgrade, Serbia}

\centerline{$^1$E-mail: dugic@knez.uis.kg.ac.yu}

\centerline{$^2$E-mail: mcirkovic@aob.aob.bg.ac.yu}

\bigskip

{\bf Abstract:} We investigate the rate of operation of quantum
"black boxes" ("oracles") and point out the possibility of
performing an operation by a quantum "oracle" whose average energy
equals zero. This counterintuitive result not only presents a
generalization of the recent results of Margolus and Levitin, but
might also sharpen the conceptual distinction between the
"classical" and the "quantum" information.

\bigskip

{\bf PACS:} 03.67.Lx, 03.65.Ud, 03.65.Yz

\bigskip

{\bf 1. Introduction: information-processing bounds}

\bigskip

In the realm of computation, one of the central questions is "what
limits the laws of physics place on the power of computers?"
[1,2]. The question is of great relevance to a wide range of
scientific disciplines, from cosmology and nascent discipline of
physical eschatology [3-5] to biophysics and cognitive sciences
which study information processing in the conscious mind [6, 7].
One of the physical aspects of this question refers to the minimum
time needed for execution of the logical operations, i.e.\ to the
maximum rate of transformation of state of a physical system
implementing the operation. From the fundamental point of view,
this question tackles the yet-to-be-understood relation between
the energy (of a system implementing the computation) on the one,
and the concept of information, on the other side. Apart from
rather obvious practical interest stemming from the explosive
development of information technologies (expressed, for instance,
in the celebrated Moore's law), this trend of merging physics and
information theory seems bound to offer new insights into the
traditional puzzles of physics. Specifically, answering the
question above might shed new light on the standard "paradoxes"
of the quantum world [8-10].

Of special interest are the rates of the {\it reversible
operations} (i.e.\ of the reversible quantum state
transformations). To this end, the two bounds for the so-called
"orthogonal transformations (OT)" are known; by OT we mean a
transformation of a (initial) state $\vert \Psi_i \rangle$ to a
(final) state $\vert \Psi_f \rangle$, while $\langle \Psi_i \vert
\Psi_f \rangle = 0$. First, the minimum time needed for OT can be
characterized in terms of the spread in energy, $\Delta \hat H$,
of the system implementing the transformation [11-15]. However,
recently, Margolus and Levitin [16, 17] have extended this result
to show that a quantum system with average energy $\langle \hat H
\rangle$ takes time at least $\tau = h / 4 (\langle \hat H \rangle
- E_0)$ to evolve to an orthogonal state, where $E_0$ is the
ground state energy. In a sense, the second bound is more
restrictive: a system with the zero energy (i.e. in the ground
state) {\it cannot} perform a computation ever. This however stems
nothing about the nonorthogonal evolution which is still of
interest in quantum computation.

Actually, most of the efficient quantum algorithms [18-21] employ
the so-called quantum "oracles" (quantum "black boxes") not
requiring orthogonality of the initial and the final states of the
composite quantum system "input register + output register
($I+O$)" [22-24]. Rather, orthogonality of the final states of the
subsystems' (e.g. $O$'s) states is required, thus emphasizing a
need for a new bound for the operation considered.

In this paper we show that the relative maximum of the rate of
operation of the quantum "oracles" may point out the {\it zero
average energy} of interaction in the composite system $I+O$.
Actually, it appears that the rate of an operation {\it cannot} be
characterized in terms of the average energy of the composite
system as a whole. Rather, it can be characterized in terms of
the average energy of interaction Hamiltonian. Interestingly
enough, the ground state energy $E_0$ {\it plays no role}, here,
and the absolute value of the average energy of interaction
($\vert \langle \hat H_{\rm int} \rangle \vert$) plays the role
analogous to the role of the difference $\langle \hat H_{\rm
int}\rangle - E_0$ in the considerations of OT. Physically, we
obtain: the lower the average energy, the higher the rate of
operation. This result is in obvious contradistinction with the
result of Margolus and Levitin [16, 17]---in terms of the
Margolus-Levitin theorem, our result would read: the lower the
difference $\langle \hat H_{\rm int} \rangle - E_0$, the {\it
higher\/} the rate of (nonorthogonal) transformation. On the
other side, our result is not reducible to the previously obtained
bound characterized in terms of the spread in energy [11-15], thus
providing us with a new bound in the quantum information theory.

\bigskip

{\bf 2. The quantum "oracle" operation}

\bigskip

We concern ourselves with the bounds characterizing the rate of
(or, equivalently, the minimum time needed for) the {\it
reversible} transformations of a quantum system's states.
Therefore, the bounds known for the irreversible transformations
are of no use here. Still, it is a plausible statement that the
information processing should be faster for a system with higher
(average) energy, even if---as it is the case in the classical
reversible information processing---the system does not dissipate
energy (e.g.\ [25]). This intuition of the classical information
theory is justified by  the Margolus-Levitin bound [16, 17].
However, this bound refers to OT, and does not necessarily
applies to the nonorthogonal evolution.

The typical nonorthogonal transformations in the quantum
computing theory are operations of the quantum "oracles" {\it
employing} {\it "quantum entanglement"} [18-21, 23, 24]. Actually,
the operation considered is defined by the following state
transformation:
$$\vert \Psi_i \rangle_{IO} = \sum_x C_x \vert x \rangle_I \otimes
\vert 0 \rangle_O \to \vert \Psi_f \rangle_{IO} = \sum_x C_x
\vert x \rangle_I \otimes \vert f(x) \rangle_O, \eqno (1)$$

\noindent where $\{\vert x \rangle_I\}$ represents the
"computational basis" of the input register, while $\vert 0
\rangle_O$ represents an initial state of the output register; by
"$f$" we denote the oracle transformation.

The key point is that the transformation (1) does {\it not}
[18-21] require the orthogonality condition $_{IO}\langle \Psi_i
\vert \Psi_f \rangle_{IO} = 0$ to be fulfilled. {\it Rather},
orthogonality for the {\it subsystem's states} is required
[18-21]:
$$_O\langle f(x) \vert f(x') \rangle_O = 0, \; x \neq x'\eqno (2)$$

\noindent for at least some pairs $(x, x')$, which, in turn, is
neither necessary nor a sufficient condition for the
orthogonality $_{IO}\langle \Psi_i \vert \Psi_f \rangle_{IO} = 0$
to be fulfilled.

Physical implementation of the quantum oracles of the type Eq.
(1) is an open question of the quantum computation theory. However
(and in analogy with the quantum measurement and the decoherence
process [26-31]), it is well understood that the implementation
should rely on (at least indirect, or externally controlled) {\it
interaction} in the system $I+O$ as presented by the following
equality:
$$\vert \Psi_f \rangle_{IO} \equiv \hat U(t) \vert \Psi_i \rangle_{IO}
\equiv \hat U(t) \sum_x C_x \vert x \rangle_I \otimes \vert 0
\rangle_O = \sum_x C_x \vert x \rangle_I \otimes \vert f(x, t)
\rangle_O, \eqno (3)$$

\noindent where $\hat U(t)$ represents the unitary operator of
evolution in time (Schrodinger equation) for the combined system
$I+O$; index $t$ represents an instant of time. Therefore, the
operation (1) requires the orthogonality:
$$_O\langle f(x,t) \vert f(x',t) \rangle_O = 0, \eqno (4)$$

\noindent which substitutes the equality (2).

So, {\it our task} reads: by the use of Eq.\ (4), we investigate
the minimum time needed for  establishing of the entanglement
present on the r.h.s.\ of Eq.\ (1), i.e.\ of Eq.\ (3).

\bigskip

{\bf 3. The optimal bound for the quantum oracle operation}

\bigskip

In this Section we derive the optimal bound for the minimum time
needed for execution of the transformation (1), i.e. (3), as
distinguished by the expression (4). We consider the composite
system "input register + output register ($I+O$)" defined by the
{\it effective} Hamiltonian:
$$\hat H = \hat H_I + \hat H_O + \hat H_{\rm int} \eqno (5)$$

\noindent where the last term on the r.h.s.\ of (5) represents the
interaction Hamiltonian. For simplicity, we introduce the
following assumptions: (i) $\partial \hat H / \partial t = 0$,
(ii) $[\hat H_I, \hat H_{\rm int}] = 0$, $[\hat H_O, \hat H_{\rm
int}] = 0$, and (iii) $\hat H_{\rm int} = C \hat A_I \otimes \hat
B_O$, where $\hat A_I$ and $\hat B_O$ represent unspecified
observables of the input and of the output register,
respectively, while the constant $C$ represents the coupling
constant. As elaborated in Appendix I, the simplifications
(i)-(iii) are not very restrictive. For instance, concerning point
(i)---widely used in the decoherence theory---one can naturally
relax this condition to account for the wide class of the time
dependent Hamiltonians; cf.\ Eq.\ (I.1) of Appendix I. Another way
of relaxing the condition (i) is to assume the {\it sudden
switching\/} the interaction in the system {\it on\/} and {\it
off}.

\medskip

{\bf 3.1 The bound derivation}

\medskip

Given the above simplifications (i)-(iii), the unitary operator
$\hat U(t)$ (cf.\ Eq.\ (3)) spectral form reads:
$$\hat U(t) = \sum_{x,i} \exp \{-\imath t (\epsilon_x + E_i +
C\gamma_{xi})/\hbar\} \hat P_{Ix} \otimes \hat \Pi_{Oi}. \eqno
(6)$$

The quantities in Eq. (6) are defined by the following spectral
forms: $\hat H_I = \sum_x \epsilon_x \hat P_{Ix}$, $\hat H_O =
\sum_i E_i \hat \Pi_{Oi}$, and $\hat H_{\rm int} = C \sum_{x,i}
\gamma_{xi} \hat P_{Ix} \otimes \hat \Pi_{Oi}$; bearing in mind
that $\hat A_I = \sum_x a_x \hat P_{Ix}$ and $\hat B_O = \sum_i
b_i \hat \Pi_{Oi}$, the eigenvalues $\gamma_{xi} = a_x b_i$.

>From now on, we take the system's zero of energy at the ground
state by the exchange $E_{xi} \to E_{xi} - {\bf E}_{\circ}$;
$E_{xi} \equiv \epsilon_x + E_i + C \gamma_{xi}$, ${\bf
E}_{\circ}$ is the minimum energy of the composite system---which
Margolus and Levitin [16, 17], as well as Lloyd [1], have used.
Then one easily obtains for the output-register's states:
$$\vert f(x,t) \rangle_O = \sum_i
\exp\{-\imath t (\epsilon_x + E_i + C \gamma_{xi} - {\bf
E}_{\circ})/\hbar\} \hat \Pi_{Oi} \vert 0 \rangle_O. \eqno (7)$$

Substitution of Eq.\ (7) into Eq.\ (4) directly gives:
$$D_{xx'}(t) \equiv _O\langle f(x,t) \vert f(x',t)
\rangle_O = \exp\{-\imath t (\epsilon_x - \epsilon_{x'})/ \hbar\}
\times$$
$$\times \sum_i p_i \exp\{-\imath C t (a_x - a_{x'}) b_i/
\hbar\} = 0, \quad \sum_i p_i = 1, \eqno (8)$$

\noindent where $p_i \equiv _O\langle 0 \vert \Pi_{Oi} \vert 0
\rangle_O$. The expression (8) represents the condition for
"orthogonal evolution" of subsystem's ($O$'s) states bearing
explicit time dependence; the ground energy ${\bf E}_{\circ}$
{\it does not appear} in (8).

But this expression is already known from, e.g., the decoherence
theory [26-31]. Actually, one may write:
$$D_{xx'}(t) = \exp\{-\imath t (\epsilon_x - \epsilon_{x'})
/\hbar\} z_{xx'}(t), \eqno (9)$$

\noindent where
$$z_{xx'}(t) \equiv \sum_i p_i \exp\{-\imath C t (a_x - a_{x'})
b_i/ \hbar\} \eqno (10)$$

\noindent represents the so-called {\it "correlation amplitude"},
which appears in the off-diagonal elements of the (sub)system's
($O$'s) density matrix [26]:
$$\rho_{Oxx'}(t) = C_x C_{x'}^{\ast} z_{xx'}(t).$$

So, we could make a direct application of the general results of
the decoherence theory. However, our aim is to estimate the
minimum time for which $D_{xx'}(t)$ may approach zero, rather than
calling for the {\it qualitative} limit of the decoherence theory
[26]:
$$\lim_{t \to \infty} \vert z_{xx'}(t)\vert = 0, \eqno (11)$$

\noindent or equivalently $\lim_{t \to \infty}z_{xx'}(t) \to 0$.

In order to obtain the more elaborate {\it quantitative} results,
we shall use the inequality $\cos x \ge 1 - (2/\pi) (x + \sin x)$,
{\it valid only for} $x \ge 0$ [16, 17]. However, the use cannot
be straightforward.

Namely, the exponent in the "correlation amplitude" is
proportional to:
$$(a_x - a_{x'}) b_i, \eqno (12)$$

\noindent which need not be strictly positive. That is, for a
fixed term $a_x - a_{x'} > 0$, the expression Eq. (12) can be
both positive or negative, depending on the eigenvalues $b_i$.
For this reason, we will refer to the general case of eigenvalues
of the observable $\hat B_O$, $\{b_i, -\beta_j\}$, where both
$b_i, \beta_j > 0, \forall{i, j}$.

In general, Eq. (10) reads:
$$z_{xx'}(t) = z^{(1)}_{xx'}(t) + z_{xx'}^{(2)}(t),
\eqno(13a)$$

\noindent where
$$z_{xx'}^{(1)} = \sum_i p_i \exp\{-\imath C t (a_x -
a_{x'}) b_i/\hbar\}, \eqno (13b)$$
$$z_{xx'}^{(2)} = \sum_j p'_j \exp\{\imath C t (a_x -
a_{x'}) \beta_j/\hbar\}, \eqno (13c)$$

\noindent while $\, \sum_i p_i + \sum_j p'_j = 1$. Since both
$(a_x - a_{x'}) b_i >0$, $(a_x - a_{x'}) \beta_j >0  \; \;
(\forall{i, j})$, one may apply the inequality mentioned above.

Relaxed equality (4)---or relaxed equality (11)---is equivalent to
${\rm Re} \, z_{xx'}$ $\cong 0$ {\it and} ${\rm Im} \, z_{xx'}
\cong 0$. Now, from Eq.\ (13a-c) it directly follows:
$${\rm Re} \, z_{xx'} = \sum_i p_i \cos [C (a_x - a_{x'}) b_i t /
\hbar]$$

$$+ \sum_j p'_j \cos [C (a_x - a_{x'}) \beta_j t / \hbar], \eqno
(14)$$

\noindent which, after applying the above inequality gives:
$${\rm Re} \, z_{xx'} > 1 - {4 \over h} C (a_x - a_{x'})
(B_1 + B_2) t - {2 \over \pi} {\rm Im} \, z_{xx'} -$$
$$- {4 \over \pi} \sum_i p_i \sin [C (a_x - a_{x'}) b_i
t/\hbar], \eqno (15)$$ \noindent where $B_1 \equiv \sum_i p_i
b_i$, and $B_2 \equiv \sum_j p'_j \beta_j$.

Since $\vert \sum_i p_i \sin [C (a_x - a_{x'}) b_i t/\hbar]\vert
\le \sum_i p_i \equiv \alpha < 1, \quad \forall{t}$, from Eqs.\
(11) and (15) it follows:
$$0 \cong {\rm Re} \, z_{xx'} + {2 \over \pi} {\rm Im} \, z_{xx'} >
1 - {4 \alpha\over \pi} - {4 \over h} C (a_x - a_{x'}) (B_1 +
B_2) t. \eqno(16)$$

>From (16) it is obvious that the condition imposed by Eq.\ (4)
cannot be fulfilled in time intervals shorter than $\tau_{xx'}$:
$$\tau_{xx'} > {(1 - 4 \alpha / \pi) h \over
4 C (a_x - a_{x'})(B_1 + B_2)}. \eqno (17)$$

\noindent The expression is strictly positive for $\alpha <
\pi/4$, and which directly defines the optimal bound $\tau_{\rm
ent}$ as:
$$\tau_{\rm ent} = {\rm sup} \, \{\tau_{xx'}\}. \eqno (18)$$

The assumption $\alpha < \pi / 4$ is not very restrictive.
Actually, above, we have supposed that neither $\sum_i p_i \cong
1$, nor $\sum_j p'_j \cong 1$, while the former is automatically
satisfied with the condition $\alpha < \pi/4$.

\medskip

{\bf 3.2 Analysis of the results}

\medskip

The bound $\tau_{\rm ent}$ is obviously determined by the minimum
difference $a_x - a_{x'}$. This difference is virtually
irrelevant (in the quantum computation models, it is typically of
the order of $\hbar$). So, one may note, that the bound in Eq.\
(18) may be {\it operationally} decreased by the increase in the
coupling constant $C$ and/or by the increase in the sum $B_1 +
B_2$. As to the former, for certain quantum "hardware" [32, 33],
the coupling constant $C$ may be partially manipulated by
experimenter. On the other side, similarly---as it directly
follows from the above definitions of $B_1$ and $B_2$---by the
choice of the initial state of the output register, one could
eventually increase the rate of the operation by increasing the
sum $B_1 + B_2$.

Bearing in mind the obvious equality:
$$\langle \hat H_{\rm int}\rangle = C \langle \hat A_I \rangle
\langle \hat B_O \rangle =  C \langle \hat A_I \rangle (B_1 -
B_2), \eqno (19)$$

\noindent one directly concludes that adding energy to the
composite system as a whole, does not necessarily increase the
rate of the operation considered. Rather, the rate of the
operation is determined by the {\it absolute value of the average
energy of interaction}, $\vert \langle \hat H_{\rm int} \rangle
\vert$. For instance, if $B_1 \neq 0$ while $B_2 = 0$ (or $B_2
\neq 0, B_1 = 0$), from Eq. (19) it follows that the increase in
$B_1$ (or in $B_2$, and/or in the coupling constant $C$) coincides
(for $\langle \hat A_I \rangle \neq 0$) with the increase in
$\vert \langle \hat H_{\rm int} \rangle\vert$, as well as with the
decrease in the bound Eq. (18). This observation is in accordance
with the Margolus-Levitin bound [16, 17]: the increase in the
average energy (of interaction) gives rise to the increase in the
rate of the operation (still, without any restrictions posed by
the minimum energy of either the total, or the interaction
Hamiltonian). Therefore, the absolute value $\vert \langle \hat
H_{\rm int} \rangle\vert$ {\it plays, in our considerations, the
role analogous} to the role of the difference $\langle \hat
H_{\rm int} \rangle - E_0$ in the considerations of the
"orthogonal transformations".

However, for the general initial state of the output register,
both $B_1 \neq 0$ and $B_2 \neq 0$. Then, e.g., for $B_1 > B_2$:
$$B_1 + B_2 = B_1 (1 + \kappa) \le 2B_1, \; \kappa \le 1, \eqno (20)$$

\noindent which obviously determines the {\it relative maximum\/}
of the rate of the operation by the following equality:
$$B_1 = B_2, \; \kappa = 1, \eqno (21a)$$

\noindent which, in turn (for $\langle \hat A_I \rangle \neq 0)$,
is equivalent with:
$$\langle \hat H_{\rm int} \rangle = 0. \eqno (21b)$$

But this result is in {\it obvious contradistinction\/} with the
result of Margolus and Levitin [16, 17]. Actually, the expressions
(21a,b) stem that, apart from the concrete values of $B_1$ and
$B_2$, the relative maximum of the rate of the operation {\it
requires} (mathematically: implies) {\it the zero average energy
of interaction}, $\langle \hat H_{\rm int} \rangle = 0$---which
(as distinguished above) is analogous to the equality $\langle
\hat H_{\rm int} \rangle - E_0 = 0$ for "orthogonal
transformations."

\bigskip

{\bf 4. Discussion}

\bigskip

Intuitively, the speed of change of a system's state should be
directly proportional to the average energy of the system. This
intuition is directly justified for the quantum OT by the
Margolus-Levitin theorem [16, 17]. Naively, one would expect this
statement to be of relevance also for the nonorthogonal
evolution. Actually, in the course of the orthogonal evolution,
the system's state "passes" through a "set" of nonorthogonal
states, thus making nonorthogonal evolution faster than the
orthogonal evolution itself. (At least this physical picture is
justified for "realistic" interpretations of quantum mechanics,
like the dynamical-reduction or many-worlds theories.)

This intuition is obviously incorrect for the cases studied. In a
sense, the expressions (21) state the opposite: the lower the
difference $B_1 - B_2$ (i.e. the lower the average energy of
interaction, $\vert \langle \hat H_{\rm int} \rangle \vert$), the
faster the operation considered. Therefore, our main result, Eq.
(21), is in obvious contradistinction with the conclusion drawn
from the Margolus-Levitin bound [16, 17]: the {\it zero energy
quantum information processing is possible\/} and, in the sense of
Eq. (21), is {\it even preferable}. From the {\it operational}
point of view, the bound $\tau_{\rm ent}$ can be decreased by
manipulating the interaction in the combined system $I+O$, as
well as by the proper {\it local operations\/} (e.g., the proper
state preparations increasing the sum $B_1 + B_2$) performed on
the output register.

As it can be easily shown, the increase in the sum $B_1 + B_2$
coincides with the increase in the spread in $\hat B_O$, $\Delta
\hat B_O$, i.e. with the increase in the spread $\Delta \hat
H_{\rm int}$. This observation however cannot be interpreted as to
suggest reducibility of the bound in Eq.\ (17) onto the bound
characterized in terms of the spread in energy [11-15]---in the
case studied, $\Delta \hat H_{\rm int}$. Actually, as it is rather
obvious, the increase in the spread $\Delta \hat H_{\rm int}$ {\it
does not} pose any restrictions on the average value $\langle
\hat H_{\rm int} \rangle$. Therefore, albeit having a common
element with the previously obtained bound [11-15], the bound in
Eqs.\ (17) and (18), represents a new bound in the quantum
information theory.\footnote{$^*$}{This bound is of interest also
for the decoherence theory, but it does not provide us with
magnitude of the "decoherence time", $\tau_D$. Actually, one may
write---in our notation---that $\tau_D \propto (a_x -
a_{x'})^{-2}$, while---cf.\ Eq. (17)---$\tau_{\rm ent} \propto
(a_x - a_{x'})^{-1}$, which therefore indicates $\tau_D \gg
\tau_{\rm ent}$. This relation is in accordance with the general
results of the decoherence theory: the entanglement formation
should precede the decoherence effect.}

From Eq.\ (17), one directly determines the absolute maximum of
the rate of the operation, i.e.\ the absolute minimum of the
r.h.s.\ of Eq.\ (17). Actually, for $\hat B_{O}$ bounded (which is
generally the case for quantum computation models), the
inequality $B_{\rm min} \le B_1 + B_2 \le B_{\rm max}$ determines
the absolute minimum of the r.h.s.\ of Eq.\ (17):
$${(1 - 4 \alpha/\pi) h \over 4C (a_x - a_{x'}) B_{\rm max}},
\eqno(22)$$

\noindent where $B_{\rm max}$ ($B_{\rm min}$) is the maximum
(minimum) in the set $\{b_i, \beta_j\}$. Interestingly enough,
the minimum Eq.\ (22) is achievable also in the following special
case: if $B_1 + B_2 \equiv p_1 b_{\rm max} + p'_1 \beta_{\rm
max}$, while $p_1 = p'_1 = 1/2$ and $b_{\rm max} = \beta_{\rm max}
\equiv B_{\rm max}$, one obtains, again, that $\langle \hat
H_{\rm int} \rangle = 0$; by $b_{\rm max}$ ($\beta_{\rm max}$) we
denote the maximum in the set $\{b_i\}$ ($\{\beta_j\}$).

It cannot be overemphasized: the zero (average) energy quantum
information processing is in principle possible. Moreover, the
condition $\langle \hat H_{\rm int} \rangle = 0$ determines the
relative maximum of the operation considered. But this result
challenges our classical intuition, because it is commonly
believed that the efficient information processing presumes an
"energy cost". In the classical domain, this was established in
1960s by Brillouin [25], following the ground-breaking studies of
Szilard and others on the problem of Maxwell's demon. So, one may
wonder if "saving energy" might allow the efficient information
processing {\it ever}. Without ambition to give a definite answer
to this question, we want to stress: as long as the "energy cost"
in the classical information processing (including the
quantum-mechanical "orthogonal evolution") is surely necessary,
this {\it need not be the case with the quantum information
processing}, such as the entanglement establishing. Actually, the
"classical information" refers to the orthogonal (mutually
distinguishable) states, while dealing exclusively with the
orthogonal states is a basis of the classical information
processing [23]. {\it Nonorthogonal} states (i.e. nonorthogonal
transformations) we are concerned with necessarily refer to the
{\it non}classical information processing. So, without further
ado, we stress that Eq.\ (21) exhibits a peculiar aspect of the
quantum information (here: of the entanglement formation), so
pointing to the necessity of its closer further study.

The roles of the two registers ($I$ and $O$) are by definition
asymmetric, as obvious from Eqs.\ (1) and (3). This asymmetry is
apparent also in the bound given in Eq.\ (17), which is the reason
we do not discuss in detail the role of the average value $\langle
\hat A_I \rangle$. Having in mind the considerations of Section
3, this discussion is really an easy task not significantly
changing the conclusions above.

Finally, {\it applicability of the bound\/} (17) for the general
purposes of the quantum computing theory is limited by the
defining expression Eq.\ (1). The bound in Eq.\ (17) {\it is of no
use\/} for the algorithms {\it not\/} employing quantum
entanglement. As such an example, we may consider Grover's
algorithm [34], which does not employ quantum entanglement in its
oracle operation. As another example, we mention the so-called
"adiabatic quantum computation" (AQC) model [35, 36]. This new
computation model does not employ any "oracles" whatsoever.
Moreover, the AQC algorithms typically involve non-persistent
entanglement (as distinct from those in Eq. (1)) of states of
neighbor qubits (cf.\ Eq.\ (II.1) in Appendix II). Therefore, the
bound in Eq.\ (17) is of no direct use in AQC, and cannot be used
for analyzing this {\it non-circuit model\/} of quantum
computation (cf.\ Appendix II).

The work on application of Eq.\ (17) in optimizing entangling
circuits is in progress, and will be published elsewhere.

\bigskip

{\bf 5. Conclusion}

\bigskip

We show that the zero average energy quantum information
processing is theoretically possible. Specifically, we show that
the entanglement establishing in the course of operation of some
typical quantum oracles employed in the quantum computation
algorithms distinguishes the zero average energy of interaction
in the composite system "input register + output register". This
result challenges our classical intuition, which plausibly stems
a need for the "energy cost" in the information processing. To
this end, our result, which sets a new bound for the nonorthogonal
evolution in the quantum information processing, sets a new
quantitative relation between the concept of information on the
one, and of the physical concept of energy, on the other
side---the relation yet to be properly understood.
\bigskip

{\bf Literature:}

\item{[1]}
S.\ Lloyd, {\it Nature\/} {\bf 406}, 1047 (2000).

\item{[2]} A. Galindo and M. A. Martin-Delgado, {\it Rev. Mod.
Phys.}~{\bf 74}, 347 (2002).

\item{[3]} S.\ Lloyd, {\it Phys. Rev. Lett.} {\bf 88}, 237901-1 (2002)

\item{[4]} F. J. Tipler, {\it Int. J. Theor. Phys.} {\bf 25}, 617-661
(1986).

\item{[5]} F. C. Adams and G. Laughlin, {\it Rev. Mod. Phys.} {\bf 69},
337-372 (1997).

\item{[6]} C. H. Woo, {\it Found. Phys.} {\bf 11}, 933 (1981).

\item{[7]} S. Hagan, S. R. Hameroff, and J. A. Tuszynski, {\it
Phys. Rev. E} {\bf 65}, 061901 (2002).

\item{[8]}
B.\ d'Espagnat, "Conceptual Foundations of Quantum Mechanics"
(Benjamin, Reading, MA, 1971).

\item{[9]}
J.\ A.\ Wheeler and W.\ H. Zurek, (eds.), "Quantum Theory and
Measurement" (Princeton University Press, Princeton, 1982).

\item{[10]}
Cvitanovi\'c et al, (eds.), "Quantum Chaos--Quantum Measurement"
(Kluwer Academic Publishers, Dordrecht, 1992).

\item{[11]}
S.\ Braunstein, C. Caves and G.\ Milburn, {\it Ann. Phys.} {\bf
247}, 135 (1996).

\item{[12]}
L. Mandelstam and I. Tamm, {\it J. Phys.} (USSR) {\bf 9}, 249
(1945).

\item{[13]}
A. Peres, "Quantum Theory: Concepts and Methods" (Kluwer Academic
Publishers, Hingham, MA, 1995).

\item{[14]}
P. Pfeifer, {\it Phys. Rev. Lett.} {\bf 70}, 3365 (1993).

\item{[15]}
L. Vaidman, {\it Am. J. Phys.} {\bf 60}, 182 (1992).

\item{[16]}
N. Margolus and L.\ B. Levitin, in {\it Phys. Comp96}, (eds.) T.
Toffoli, M. Biafore and J. Leao (NECSI, Boston 1996).

\item{[17]}
N. Margolus, L.\ B. Levitin, {\it Physica D} {\bf 120}, 188
(1998).

\item{[18]}
D.R. Simon, {\it SIAM J. Comp.} {\bf 26}, 1474 (1997).

\item{[19]}
P.W. Shor, {\it SIAM J. Comp.} {\bf 26}, 1484 (1997).

\item{[20]}
Peter W. Shor, "Introduction to Quantum Algorithms", e-print
arXiv quant-ph/0005003.

\item{[21]}
M. Ohya and N. Masuda, {\it Open Sys. \& Information Dyn.} {\bf
7}, 33 (2000).

\item{[22]}
A.M. Steane, {\it Rep. Prog. Phys.} {\bf 61}, 117 (1998).

\item{[23]}
M. Nielsen and I. Chuang, "Quantum Information and Quantum
Computation" (Cambridge University Press, Cambridge, 2000).

\item{[24]}
J.Preskill, in "Introduction to Quantum Computation and
Information", (eds.) H.K. Lo, S. Popescu, J. Spiller, World
Scientific, Singapore, 1998.

\item{[25]} L.\ Brillouin, "Science and Information Theory" (New York:
Academic Press, 1962).

\item{[26]}
W.H. Zurek, {\it Phys. Rev. D\/} {\bf 26}, 1862 (1982).

\item{[27]}
W.H. Zurek, {\it Prog. Theor. Phys.} {\bf 89}, 281 (1993).

\item{[28]}
D. Giulini, E. Joos, C. Kiefer, J. Kupsch, I.-O. Stamatescu and
H.D. Zeh, "Decoherence and the Appearance of a Classical World in
Quantum Theory" (Springer, Berlin, 1996).

\item{[29]}
M. Dugi\' c, {\it Physica Scripta\/} {\bf 53}, 9 (1996).

\item{[30]}
M. Dugi\' c, {\it Physica Scripta\/} {\bf 56}, 560 (1997).

\item{[31]}
W.H. Zurek, {\it Phys. Today\/} {\bf 48}, 36 (1991).

\item{[32]}
Fortschritte der Physik {\bf 48}, Issue 9-11 (2000).

\item{[33]}
Quantum Information \& Computation, {\bf 1}, Special Issue,
December, 2001

\item{[34]}
L. Grover, {\it Phys. Rev. Lett.} {\bf 78}, 325 (1997).

\item{[35]}
E. Farhi et al, "Quantum computation by adiabatic evolution",
e-print arXiv quant-ph/0001106.

\item{[36]}
E. Farhi et al, {\it Science\/} {\bf 292}, 472 (2001).

\bigskip

{\bf Appendix I}

\bigskip

Relaxing the simplifications (i)-(iii) of Section 3 does not lead
to significant changes of our results. This can be seen by
employing the arguments of Dugi\' c [29, 30], but for
completeness, we briefly outline the main points in this regard.

First, for a time dependent Hamiltonian, which is still a {\it
"nondemolition observable"}, $[\hat H(t), \hat H(t')] = 0$, the
spectral form [30]:
$$\hat H = \sum_{x,i} \gamma_{xi}(t) \hat P_{Ix} \otimes
\hat \Pi_{Oi}. \eqno (I.1)$$

\noindent This is a straightforward generalization of the cases
studied and also a wide class of the time dependent Hamiltonians.
E.g., from (I.1) it easily follows that the term $\alpha_{xi}(t) =
\int\limits_0^t \gamma_{xi}(t') dt'$ substitutes the term
$\gamma_{xi} t$ in the exponent of the expression (6). Needless
to say, this relaxes the constraint (i) of Section 3 and makes
the link to the realistic models of the quantum "hardware".

To this end, it is worth emphasizing: in the realistic models one
assumes the actions performed on the qubits in order to design
the system dynamics. Interestingly enough, such actions usually
result in the (effective) {\it time independent\/} model
Hamiltonians [23, 33]. Moreover, some time dependent models allow
direct applicability of the above notion; e.g., for the
controlled Heisenberg interaction, $\hat H = J(t) \vec S_1 \cdot
\vec S_2$, the action reads: $\hat U(t) = \exp (- \imath K(t)
\vec S_1 \cdot \vec S_2)$, where $K(t) \equiv \int\limits_0^t
J(t') dt'$. As an illustration of the models not fitting (i), one
can consider NMR models which, in turn, are known to be of only
limited use in the large-scale quantum computation [23, 32, 33].
{\it We conclude that the realistic models of the large-scale
quantum computation fit with the relaxed point (i) of our
considerations}.

Similarly, relaxing the exact compatibilities (cf.\ the point (ii)
in Section 3) leads to the approximate separability---i.e., in Eq.
(I.1), there appear terms of small norm---which does not change
the results concerning the "correlation amplitude" $z_{xx'}(t)$
[26], and consequently concerning $D_{xx'}(t)$.

Finally, generalization of the form of the interaction
Hamiltonian (cf.\ point (iii) of Section 3) does not produce any
particular problems, as long as the Hamiltonian is of (at least
approximately) separable kind, and also a "nondemolition
observable". E.g., from the {\it general form\/} for $\hat H_{\rm
int}$ [30], $\hat H_{\rm int} = \sum_k C_k \hat A_{Ik} \otimes
\hat B_{Ok}$, one obtains the term $\sum_k C_k (a_{kx} - a_{kx'})
b_{ki}$, instead of the term of Eq.\ (12).

The changes of the results may occur [30] if the Hamiltonian of
the composite system is not of the separable kind and/or not a
"nondemolition observable"; for an example see Appendix II.

For completeness, we notice: a composite-system observable is of
the {\it separable kind\/} if it can be proved {\it
diagonalizable in a noncorrelated (the tensor-product) orthonormal
basis\/} of the Hilbert space of the composite system [30].

\bigskip

{\bf Appendix II}

\bigskip

By "nonpersistent entanglement" we mean the states of a composite
system which can be written as:
$$\vert \Psi \rangle = \sum_i C_{it} \vert i_t \rangle \vert i_t
\rangle, \eqno (II.1)$$

\noindent i.e.\ states whose Schmidt (canonical) form is labeled
by an instant of time, $t$ (continuously varying with time). The
occurrence of such forms for AQC can be easily proved by the use
of the method developed in Ref. [30] applied to, e.g., Eq.\ (3.5)
of Ref. [35]. Needless to say, states of the Eq.(II.1)-form are
exactly what should be avoided in the situations described by Eq.
(1).

To this end, the problem addressed in Ref. [30] reads: "what
characteristics of the system Hamiltonian are required in order
to attain the persistent entanglement (cf.\ Eq.\ (1))?". The
answer is given by the points (i)-(iii) (but see Appendix I). In
other words, as long as the conditions (i)-(iii) are fulfilled,
nonpersistent entanglements do not occur in the system. As a
corollary, having (i)-(iii) in mind, the nonpersistent
entanglement of AQC cannot be (at least not directly) addressed
within the present considerations.

\vfill\eject

\end